\documentclass[manuscript,screen]{acmart}
\AtBeginDocument{%
  \providecommand\BibTeX{{%
    \normalfont B\kern-0.5em{\scshape i\kern-0.25em b}\kern-0.8em\TeX}}}

\newcommand*\classicon[1]{\includegraphics[height=8pt]{images/classes/#1.png} \textit{#1}}
\newcommand*\classiconType[2]{\includegraphics[height=8pt]{images/classes/#2.png} \textit{#1 #2}}
\newcommand*\classiconOnly[1]{\includegraphics[height=8pt]{images/classes/#1.png}}
\newcommand{\blue}[1]{{{#1}}}
\usepackage{multirow}
\usepackage{longtable}
\usepackage{subcaption}

\begin{document}

\title{On Video Game Balancing: Joining Player- and Data-Driven Analytics}

\author{Johannes Pfau}
\email{jopfau@ucsc.edu}
\orcid{0000-0002-8760-5023}
\author{Magy Seif El-Nasr}
\email{mseifeln@ucsc.edu}
\orcid{0000-0002-7808-1686}
\affiliation{%
  \institution{University of California, Santa Cruz}
  \streetaddress{1156 High St}
  \city{Santa Cruz}
  \state{California}
  \country{USA}
  \postcode{95064}
}
\renewcommand{\shortauthors}{Pfau and Seif El-Nasr}

\begin{abstract}
Balancing is, especially among players, a highly debated topic of video games. Whether a game is sufficiently balanced greatly influences its reception, player satisfaction, churn rates and success. Yet, conceptions about the definition of balance diverge across industry, academia and players, and different understandings of designing balance can lead to worse player experiences than actual imbalances. This work accumulates concepts of balancing video games from industry and academia and introduces a player-driven approach to optimize player experience and satisfaction. Using survey data from 680 participants and empirically recorded data of over 4 million in-game fights of Guild Wars 2, we aggregate player opinions and requirements, contrast them to the status quo and approach a democratized quantitative technique to approximate closer configurations of balance. We contribute a strategy of refining balancing notions, a methodology of tailoring balance to the actual player base and point to an exemplary artifact that realizes this process.
\end{abstract}




\maketitle

\section{Introduction \& Background}
Video game balancing is one of the most controversial topics of modern video games, especially in the context of (online) multiplayer games that undergo regular updates, both for competitive Player-versus-Player (PvP) as well as for collaborative Player-versus-Environment (PvE) games and modes. Besides the introduction of new content and bug fixes, balancing adjustments are among the driving factors of game update patches, as the often exploding complexity of choices, the diversity of player expectations and the hardly predictable impact of changes onto the actual adaption and reception of the player base render the problem of finding a well-balanced game state almost impossible. 
Games of all popular genres employ never-ending balance patch paradigms even years after the launch of a game. These can operate on various layers of in-game choices, such as the playable Champions\footnote{https://www.leagueoflegends.com/en-us/news/tags/patch-notes/} of the Multiplayer Online Battle Arena (MOBA) \textit{League of Legends} \cite{LoL}, equippable weapons\footnote{https://liquipedia.net/counterstrike/Patches} of the First-Person Shooter (FPS) \textit{Counter-Strike: Global Offensive} \cite{CSGO} or active and passive changes to classes, traits and skills\footnote{https://en-forum.guildwars2.com/forum/6-game-update-notes/} of the Massively multiplayer online role-playing game (MMORPG) \textit{Guild Wars 2} \cite{GW2}.

In contrast to ordinary bug fixes that most of the time have a well-defined optimal solution, working on balancing issues is a steady and repetitive task suffering from the unpredictability and inertia of the player base accommodating to the new game configuration (\textit{``a lack of balance [...] only becomes apparent after many months of play''})\cite{Hullett2012_EmpiricalAnalysis}, as well as from the diverging and controversial opinions and perceptions from both player base and developers (\textit{``after each patch, often the discussion begins again,
factoring in new balancing or abilities for each class''})\cite{lewis2010mining}.

What makes this process even harder to solve is that there is no clear or accepted definition of what \textit{balancing} means in the context of video games, or even what a \textit{well-balanced} state or configuration of a game constitutes.
Becker and Görlich recently stressed this incongruity, emphasized that this controversial but highly important topic is barely investigated by academia and contrasted various definitions from designers, developers and other practitioners \cite{becker2020game}. They conclude that no two of these authors share identical understandings of game balancing and even if some concepts overlap, no undisputed fundamental concepts of \textit{game balancing} protrude.

From an industrial perspective, a game might be balanced \textit{``if a reasonably large number of options available to the player are \textbf{viable} — especially, but not limited to, during high level play by expert players''} according to Sirlin \cite{sirlin2009balancing}. This however necessitates a finer distinction on when \textit{reasonably large} begins and how sharp \textit{viable} can be defined. While the former is arguably highly game- and context-related, he refers to \textit{viable options} as offering \textit{``meaningful decisions between promising alternatives''}.
A similar, yet slightly toned-down definition comes from Burgun, who asserts that game balancing is the craft of \textit{``keeping game elements relevant''} - indicating that \textit{dominating} strategies or choices (i.e., decisions that are by all means better than their alternatives) are the most harmful factors for the state of balance in a game \cite{burgun2011understanding}.
Felder approves this take by claiming that these dominating choices express \textit{``broken gameplay, which consists of strategies or even singular actions rendering a lot of decisions meaningless''}, which becomes detrimental for the overall player experience \cite{felder2015design}.

An intuitive implementation of viable options across the board would thus be \textit{\textbf{symmetry}}, i.e. the identical power or expected performance of choices. As such, players are exposed to equal starting conditions, a game becomes fair by definition and only (or predominantly) individual skill matters. While this might be desired in some contexts (and design aspects as flair or variation still can enhance symmetrical game setups), a perfectly symmetrical game bears the risk that there are no interesting choices beyond proven strategies anymore and that the player's decision actually does not matter \cite{millard2019whyAreGamesSoHard,felder2015design}. This lack of impact on the own performance could again be harmful for a player's intrinsic motivation. In that sense, even if a certain deviation from perfect symmetry creates a dominating strategy, this could be countered by implementing \textit{intransitivity} between options in competitive situations, so that single choices display strengths and weaknesses when opposed to different other choices (\textit{``while a balanced game can easily lead to a stagnation of strategy discovery, [...] slight differences in power between game elements encourage players to constantly search for new solutions against currently popular strategies''}) \cite{becker2020game, portnow2012perfect}. Provided that these can be approached with counter-strategies, Fender finds smaller power gaps as a \textit{``critical part of game balancing''} \cite{felder2015design}.

From another viewpoint, the balance of options could be assessed by their probability to lead to victory in competitive play \cite{sylvester2013designing}. Especially in these PvP scenarios, objective as well as subjective \textit{\textbf{fairness}} can be a strong indicator for the perceived balance of a game \cite{adams2014fundamentals}. Industrial approaches of these do not always draw on parameter balancing for the underlying options, but especially in competitive games, matchmaking between players of similar skill is a promising strategy to lead to 50/50 win rates \cite{DeCosterRubin2018pax}. Yet, Claypool et al. discovered that even if winning chances can be closely approximated to 50\%, such as in the highly populated matchmaking of \textit{League of Legends}, players still subjectively perceive these as unbalanced \cite{claypool2016_surrenderAt20}.

In contrast, ample scientific work highlights that \textit{\textbf{difficulty}} can be the driving factor for balancing PvE scenarios and/or single-player games \cite{adams2014fundamentals}. The field of \textit{dynamic difficulty adjustment} (DDA) -- sometimes also referred to as \textit{dynamic difficulty balancing} -- mainly pushes the understanding of balancing as the adequate (automatic) regulation of difficulty (parameters) in order to keep players within the desirable flow state between mental under- and overload \cite{csikszentmihalyi1990flow,hunicke2005case}\blue{, or ideally between ``too hard'' and ``too boring'' \cite{lomas2017difficulty}. In this respect, if perfect matches are not attainable, mental overload is still seen as producing higher enjoyment than boredom \cite{klarkowski2016operationalising}. 
}
Andrade et al. claim that \textit{``game balancing aims at providing a good level of challenge for the user''} \cite{andrade2006dynamic}, while Volz et al. keep it more general by defining the goal of the adjustment to make sure that \textit{``the resulting gameplay is as entertaining as possible''} \cite{volz2016demonstrating}. While this \textit{``modification of parameters of the constitutive and operational rules of a game''} \cite{schreiber2010game} follows technical procedures similar to the adjustment of parameters for adjusting the viability, symmetry or fairness of in-game choices or strategies, the underlying agenda and purpose considerably differs.

Apart from these dimensions, other perspectives look into the balance between \textit{``skill and luck''} \cite{schell2008art}, where skill is the main desired factor for outcome, but chance still makes up for an important mechanic in many games \cite{adams2014fundamentals}.

% ... 
Conclusively, different fields, communities and purposes announce and use different definitions of balance, the act of balancing and what it means for a game to be (well-) balanced. With this amount of discordance, open issues and even conflicting definitions and understandings, polishing video games towards maximized player satisfaction becomes apparently difficult. Even more, Brown stresses that balance has to take part on every conceivable level of a game, for \textit{``all options a game offers, including singular actions and strategies''} and maybe even more importantly, \textit{``the players’ perception of balance is just as important as the actual balance''} \cite{brown2019howGamesGetBalanced}.
This importance is elevated by Schell who claims that \textit{``to identify the right middle ground [of balancing], one has to take the audience into account''} \cite{schell2008art}, Schreiber stating that \textit{``players have differences regarding their [...] expectations''} \cite{schreiber2010game} and DeCoster and Rubin proclaiming that developers should be \textit{``open about balancing discussions to their players so they can expect solutions to existing problems''} \cite{becker2020game,DeCosterRubin2018pax}. 

The divisiveness of defining balancing notions paired with the reliance on the satisfaction of the player base for the success of the title and the convinced mindset of experts regarding player requirements strongly calls for an integrated player-driven approach of elevating game balancing, which is largely under-investigated in academia.
Even though many developers listen to professional players or communities and/or make use of recorded play data in some sense, player expectations are often not matched to balancing decisions and badly implemented update patches can lead to involuntary shifts of player behavior and/or eventual churn \cite{wang2020research,tyack2016appeal,hyeong2020whom}. 
\blue{
Prominent industrial examples of these can be found in the 1.05 patch of \textit{Uncharted 2: Among Thieves} \cite{Uncharted2} that spoiled multiplayer balance for a major part of the player base as most weapons' stats became too homogenuous (mismatching \textbf{symmetry}); \textit{Tribes: Ascend} \cite{TribesAscend} (patch 1.0.1103.1) that slowed down (and therefore eased) gameplay while the community was explicitly craving the high speed factor of the game (thus mismatching \textbf{difficulty}); or \textit{Street Fighter V} \cite{StreetFighter5} significantly losing players after a balance update (3.5) that buffed already well-performing characters and even nerfed niche fighters (diverging from \textbf{viability}).
This is only emphasized by the highly negative perceptions ($83\%$) of the \textit{Guild Wars 2} balance update in this work that raised global discontent among the community and led to a decrease of average daily playing time by $13.4\%$ when compared to the previous patch era.
}

\blue{
In sum, balancing (online) video games aims at improving the viability of different playable elements (among other factors). This often backfires when balance perceptions mismatch between developers and players, which can happen due to conflicting requirements, unawareness about how players experience game dynamics or because of the many substantially different definitions of balancing. These disagreements can lead to dissatisfaction, impaired player experience and churn, rendering the proper handling and optimization of balance a problem for the industry, games user research and human-computer interaction in general. To counteract this and fortify the understanding between developers (or researchers) and players,
}
we propose to push \textit{player-driven game balancing} to tightly couple requirements and opinions of existing player communities into the balancing process of a game.
\blue{
While this \textit{player-driven} approach aggregates the \textit{subjective} view of a player base (with all of its advantages and drawbacks), we argue that joining it with the similarly under-investigated \textit{objective} assessment of balance states in video games through \textit{data-driven} techniques can overcome its limitations by grounding opinions in empirical foundations.
This leads to the examination of the following research questions:
\begin{itemize}
    \item \textbf{(RQ1)}: With so many conflicting theoretical definitions of balancing, how can a game understand and cater to the requirements of its players? 
    \item \textbf{(RQ2)}: How can data-driven analytics help grounding the objectivity of this pool of opinions?
\end{itemize}
}

To \blue{answer these questions and assess} the feasibility of this procedure, we \blue{conducted a case study with} ($n=680$) novice up to professional players of the MMORPG \textit{Guild Wars 2}, capture their mindset towards Becker and Görlich's balance criteria of \textit{\textbf{viability}}, \textit{\textbf{symmetry}}, \textit{\textbf{fairness}} and \textit{\textbf{difficulty}}, and contrast this to the direction of actual 
balancing implementations. To validate the appropriateness of these opinions, we draw on data-driven analytics encompassing ($n_{a}=154,145$) unique player accounts and ($n_{l}=4,318,009$) atomic in-game combat logs of the game.
\\

Eventually, by delivering a grounded, systematic approach of determining detailed balance requirements of players - exemplified through an ambitious community case study that is likely to generalize to other games within and without the genre -
we contribute to the fields of player-centric design and development, games user research, game evaluation and strive to closer connect academia, industry and the very players.

\section{Related Work}
For other application cases, player-driven approaches have already delivered promising solutions that exceeded the capabilities of complete in-house implementations. 
One example would be Da Silva et al.'s utilization of the collective power of a community to retrieve a widely nested narrative and story background merely from player input \cite{DaSilva2013}.
Concerning procedural content generation, Shaker et al. highlighted the capabilities of player-centric approaches, including personalization of in-game maps or experiences \cite{Shaker2012}.
Partlan et al. utilized participatory design in order to assess requirements and develop design-driven features for co-creative game AI design tools \cite{partlan2021design}.
Even completely player-driven game development cycles have been shown to result in novel experiences, dynamic design procedures and central game features that are inherently tailored to the actual target audience \cite{Lessel2019}.
Canossa and Drachen argue that increasing the players' agency and influence on the development process can lead to enhanced experiences and immersion when introducing play-personas for customized gameplay \cite{Canossa2009} -- which arguably extends similarly to continual balancing updates.
Eventually, player-driven paradigms might even scale to the large magnitudes of communities that popular modern games accumulate, as Ma et al. indicate in their work on user innovation evaluation strategies, incorporating over 21,000 players that produced novel and sufficiently complex ideas and suggestions  \cite{Ma2019_UserInnovationCommunity}.

However, purely player-driven balancing decisions might still be warped by diverging opinions, missing empirical knowledge about the actual state of the game and the gap of experience (and requirements) between novice and expert players. The arguably most objective measures to counter lacks of knowledge and to condense how games actually play out are empirical data-driven methods \cite{GameAnalytics2013, Wallner2019}. The majority of these data-driven approaches are stemming from and/or focusing on delivering insights for the game industry (e.g. data mining, classification or prediction) \cite{Drachen2009}, targeting measures against churn \cite{Hadiji2014}, facilitating content generation \cite{Risi2019}, or easing the burden of testing \cite{Albaghajati2020} (among other areas).
Most of the remaining approaches follow academic interests concerning similar topics or fundamental (psychological or technical) regularities, structures and concepts \cite{Drachen2018, Yee2006}, often employing visualizations to gather insights for researchers or analysts \cite{bowman2012toward}.
Prominent implementations target spatio-temporal movement \cite{moura2011visualizing,wallner2019aggregated,wallner2012spatiotemporal,ahmad2019modeling}, decision making of individual or aggregate players \cite{loh2016comparison, nguyen2021glyph} or higher-level metrics and statistics \cite{drachen2012guns}.
Effectively, certain academic approaches already addressed the balancing of viable game options, such as automatic symmetric and intransitive player modeling approaches from Pfau et al. \cite{pfau2020dungeons}, asymmetric Monte-Carlo balancing from Beau and Bakkes \cite{beau2016automated}, Jaffe et al.'s maximization of fair and useful card game cards \cite{jaffe2012evaluating} or Leigh et al.'s reduction of dominant strategies through coevolution \cite{leigh2008using}. Even if most of these draw on simulation or calculation towards well-balanced game states and (to the best of our knowledge) no academic work included the players' perspective so far, it is reasonable to hypothesize that similar balancing solutions can follow or implement opinionative inputs.


Ultimately, we want to empower and harness \textit{player-driven} balancing conceptions informed by \textit{data-driven} methods. 
Making video game data transparent, explainable and applicable to its players is already one of the driving topics within the areas of game-related explainable AI and player modeling \cite{zhu2021open,lucero2020human,wells2021explainable}, and comparably holds in the context of balancing. For this reason, we developed, published and populated the player- and data-driven \textit{Guild Wars 2} analytics platform \textit{Guild Wars 2: Wingman}\footnote{https://gw2wingman.nevermindcreations.de/}, constructed and refined over a 18-month participatory development cycle 
\cite{wingman2023}.

In the following sections, we \blue{briefly introduce the environment of \textit{Guild Wars 2},} report on the acquisition and interpretation of the opinions and requirements of \blue{its} community, ground these conceptions with empirical data, and extract quantifiable factors that led to the development of a democratic player-driven balancing instrument.

\blue{
\section{Game Environment}
\label{sec:gw2}
The game used for the subsequent case study (\textit{Guild Wars 2}) is a prototypical MMORPG featuring single- and multiplayer content in storylines, open world events, various PvP modes and endgame encounters such as raid bosses, fractals or strike missions. The latter make for a large share of players' time spent in game and as they are set in fixed scenarios with only few probabilistic factors and established group compositions, strategies and roles, they enable very comparable benchmarks. Based on these, large communities formed around discussions and optimizations to overcome these challenges from which balancing discrepancies can become apparent quickly.

As \textit{Guild Wars 2} is specifically designed to not feature power creep mechanics such as increasing level caps or item qualities over time, the vast majority of players of this endgame content participates on an identical or very similar character and equipment attribute level, and performance is mainly influenced by in-game proficiency, mastery of the classes, encounter knowledge, strategy and group composition, which further adds to the comparability of the data with regards to balancing.
Still, when accepting certain degrees of noise or acknowledging these factors of variance by clustering players into equipment tiers or approximately subtracting out these confounding variables, the balancing assessment as presented here is likely to produce similarly powerful insights for instanced (group) PvE content in general, such as raids, dungeons, trials or ultimate encounters in \textit{World of Warcraft}\cite{WoW}, \textit{The Elder Scrolls Online}\cite{ESO} or \textit{Final Fantasy XIV}\cite{FFXIV}.

\textit{Guild Wars 2} features a variety of playable class options (from now on referred to as \textit{professions}), where each of the nine core professions can be expanded by particular specializations that can add further capabilities or open up new roles for this profession (e.g. the \classicon{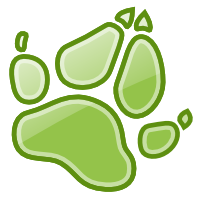} profession can be augmented with the \classicon{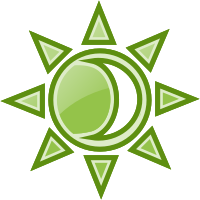} specialization to add a support dimension to the character or with the \classicon{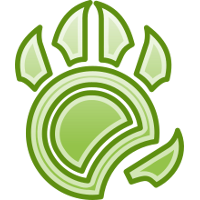} specialization to increase the damage potential). 
In theory, the professions of \textit{Guild Wars 2} allow abundant combinations of character constellations, such as individually distributed equipment attributes, chosen passive character traits and active weapon type and skill choices. In that way, players can represent and play out different roles within their party, such as dealing direct damage, damage over time, offering support or different degrees of mixtures of these factors. However, for the sake of optimization and role compression within the group, most of the time, these builds are min-maxed towards the roles of maximal damage per second (dps), full support (heal and buff application) or offensive support (dps and buff application).
The majority of players uses builds and equipment that maximizes their functionality in one of these roles with only situational variation, following community guides and recommendations.
Thus, the data-driven procedure in this paper automatically classifies recorded players into buckets of full support, offensive support, direct damage and damage over time. 
All professions are (in theory) capable of fulfilling all of these roles, yet their viability and efficiently greatly diverge (and differ with respect to the combated encounters), which constantly raises balancing gaps between builds.

For further terminology, the appendix lists explanations for all game-specific terms used across this work.
}

\section{Survey}
\label{Survey}
To sufficiently understand and represent the mindset of a player community regarding balancing, we distributed a mixed-methods survey among all major important communication channels of the affiliated game. These included the official \textit{Guild Wars 2} forums\footnote{\url{https://en-forum.guildwars2.com/}}, the /r/GuildWars2 subreddit\footnote{\url{https://www.reddit.com/r/Guildwars2/}} and several Discords and community portals of beginner/training as well as speedrun groups, such as Snow Crows\footnote{\url{https://snowcrows.com/},} Lucky Noobs\footnote{\url{https://lucky-noobs.com/}}, Discretize\footnote{\url{https://discretize.eu/}}, Hardstuck\footnote{\url{https://hardstuck.gg/}}, The Crossroads Inn and other interested communities. 
In order to capture the essence of players' stances on balance and balancing discussions, the survey followed a major game update patch \blue{(June 28, 2022)}\footnote{\url{https://wiki.guildwars2.com/wiki/Game\_updates/2022-06-28}} that impacted the viability and performance of almost all available classes and builds.
We specifically focused on the requirements to balance professions for the endgame PvE modes of the game \blue{(raids, fractals and strike missions)}, as these are highly comparable between players and groups, and we do not want to confuse potentially differing notions of balancing between PvP and PvE for this article. These instanced endgame modes are executed on the maximal level of the game, require comparable equipment and are designed on an either 5- or 10-player basis. To keep the questionnaire as understandable for the player base as possible, it frequently uses in-game terminology and game-specific notions that are explained in the appendix (\ref{appendix}) of this work for comprehension.

\subsection{Construction}
\label{sec:construction}
To transfer Becker and Görlich's concepts of balancing \cite{becker2020game} to actual in-game metrics and mechanisms of \textit{Guild Wars 2}, we composed a questionnaire of items that unravel these concepts into particular statements. Combining the authors' expertise with the help of the community around the player-driven analytics tool presented in previous work \cite{wingman2023}, these statements could be refined and successfully harmonized with in-game observations and metrics, as listed in the following section. 
Resuming the findings of this upfront discussion, a \textbf{viable} option in the challenging endgame content of \textit{Guild Wars 2} is a choice (of a profession or build) that increases the chances of a group's success and/or increases a group's efficiency (items V1-3, cf. Table \ref{tab:SurveyItems}). Increasing success is also referred to as providing \textit{utility} (e.g. heals, buffs, debuffs) for the group, while the efficiency of a profession (or player) can be closest measured by their contribution of damage per second (dps) against the boss or enemies. Thus, items of the questionnaire frequently refer to the theoretical or realistic contribution of dps or utility of a player choosing a specific profession (V4-7).
\textbf{Symmetry} could be found if all professions would deal equal amounts of dps and/or utility (S1). Opposed to that, unique buffs, i.e. utility skills or traits that are exclusively provided by specific professions, diverge from symmetry, which carried a special meaning to players in this case, as they were removed with said update patch (S2,3).
\textbf{Fairness} means that no matter which option a player chooses, they have a fair chance of achieving success. This can imply that dps values of that profession are competitive to the performances to other players, or that the chance of getting accepted in a group (and not getting kicked) might rather depend on a player's personal skill than on the profession they choose to play (F1,2). As common in online games, more advanced players and groups publish the most efficient strategies, builds and rotations to play professions and/or bosses - which might impact how players look upon these optimized builds and alternatives to these (F3,4).
Eventually, the \textbf{difficulty} of a playable profession or build depends on a variety of design decisions, game mechanics and dynamics bound to that class, which can unfold in the survivability of a class, the speed and complexity of the \textit{ideal rotation} (i.e. the most efficient skill sequence regarding individual dps; D1-3) or the difficulty of maintaining constant damage uptime throughout the various strategies against the bosses (e.g. the ability of attacking from range versus melee combat, the freedom of being able to move while executing skills versus being animation-locked, or being dependent on movement or action patterns of the bosses; D4).
Besides other questions of interest, Section \ref{Measures} outlines the specific items that were used in the actual following survey.

\subsection{Measures}
\label{Measures}
We recorded prior experience in the game as well as its main endgame content modes in years, and their satisfaction with the recent update and their agreement with the detailed balancing statements on 7-point Likert scales. 
To support the validity of our measures, we calculated the discriminant validity between those scales, which resulted in very distinct scales (cf. Table \ref{tab:SurveyItems}; overall $r_{d\_overall}=0.06$). We omitted convergent validity in this case, as we did not add similar assessment scales for the sake of survey brevity. Survey responses that consisted of only repeating (or empty) quantitative answers or showed obviously non-serious qualitative replies were excluded from the analysis.

\blue{In accordance with the game's design, the items explicitly concerned builds fully focusing on dps, as support builds are principally dps builds that only sacrifice some damage potential for utility. 
}
Above that, participants could indicate which professions needed a buff (viability increase) or nerf (viability decrease) in their opinion and were able to comment on their requirements and perceptions qualitatively. Table \ref{tab:SurveyItems} outlines all items regarding balancing opinions, attributed to the formerly introduced concepts, \blue{whereas Table \ref{tab:QualitativeItems} lists the set of qualitative questions.}

\begin{table}[!h]
    \centering
    \begin{tabular}{r|r l|r}
& \# & Item & $|r_d|$\\
\midrule
\parbox[t]{2mm}{\multirow{7}{*}{\rotatebox[origin=c]{90}{\textbf{Viability}}}}& (V1) & Every profession should be viable in endgame PvE (with at least one specialization). &\footnotesize $0.02$\\
    & (V2) &  Every specialization should be viable in endgame PvE.&\footnotesize $0.14$\\
    & (V3) &  Every profession should have a viable option for power, condition and support. &\footnotesize $0.08$\\
    & (V4) &  Power and condition builds should be equally important for endgame PvE.&\footnotesize $0.07$\\
    & (V5) &  Selfish dps builds (that do not support the squad otherwise) should exist in the game. &\footnotesize $0.12$\\
    & (V6) &  If selfish dps builds exist, they should reach higher dps than builds that contribute elsewise to the squad.&\footnotesize $0.07$ \\
    & (V7) &  Low-opportunity-cost utility and support skills should lower the overall dps output of a build. &\footnotesize $0.09$\\
    \midrule
\parbox[t]{2mm}{\multirow{3}{*}{\rotatebox[origin=c]{90}{\textbf{Symm.}}}}& (S1) & Every profession build should do equally high dps on average.&\footnotesize $0.07$\\
         &  (S2) & Unique buffs construct class identity and lead to more diversity in squad building.&\footnotesize $0.01$\\
         & (S3) & Unique buffs limit the freedom of squad formation and thus are detrimental for class diversity.&\footnotesize $0.01$\\
 \midrule
\parbox[t]{2mm}{\multirow{5}{*}{\rotatebox[origin=c]{90}{\textbf{Fairness}}}}& 
(F1) & The raid community rejects players playing classes with lower dps benchmarks.&\footnotesize $0.06$\\
    & (F2) & The fractal community rejects players playing classes with lower dps benchmarks.&\footnotesize $0.08$\\ 
    & (F3) & Speedrun guilds unnecessarily raise the expectations of the endgame PvE community with optimized
\\&&rotations and guides.&\footnotesize $0.12$\\
    & (F4) & Speedrun guilds help defining standards that makes raiding easier for new players in the first place. &\footnotesize $0.1$\\
    \midrule
\parbox[t]{2mm}{\multirow{8}{*}{\rotatebox[origin=c]{90}{\textbf{Difficulty}}}}& 
(D1) & Less complex builds and rotations should be able to perform decently (to enable beginner-friendly entry
\\&&to endgame PvE).&\footnotesize $0.02$\\
    & (D2) & Ideal execution of more complex rotations/builds should be rewarded with higher outgoing dps (than easy 
\\&&rotations/builds).&\footnotesize $0.00$\\
    & (D3) & Poor execution of more complex rotations/builds should be punished with less outgoing dps (than easy
\\&&rotations/builds).&\footnotesize $0.03$\\
    & (D4) & Easier damage uptime (e.g. by attacking from range) should be less dps-rewarding than harder-to-
\\&&optimize damage uptime (e.g. by relying on melee attacks, ground target skills or bigger hitboxes).&\footnotesize $0.04$\\

    
    \end{tabular}
    \caption{Questionnaire items of the players' understanding of balance, tailored to the use case of \textit{Guild Wars 2} PvE. All statements were answered on 7-point Likert scales from "Strongly Disagree" to "Strongly Agree" and show small discriminant validity ($|r_{d}|$).}
    \label{tab:SurveyItems}
\end{table}

\begin{table}[!h]
    \centering
    \blue{
    \begin{tabular}{|r l|}
    \midrule
            (Q1) &  What do you think are the most important factors to consider for balancing Guild Wars 2's endgame PvE?\\
            (Q2) &  How would your ideal configuration of balance look like? \\
            (Q3) &  Can you think of balancing decisions that were well-intentioned but unwelcome in the player base?\\
            (Q4) &  What do you like about the Summer Balance Patch? \\
            (Q5) &  What do you dislike about the Summer Balance Patch? \\
            (Q6) &  Do you have any additional remarks or opinions? \\
        \midrule   
    \end{tabular}
        }
    \caption{\blue{Qualitative open-ended questions asked subsequently to the questionnaire.}}
    \label{tab:QualitativeItems}
\end{table}

\subsection{Procedure}
The survey was published on the same day of the formerly mentioned balance update patch and kept open for four weeks to allow a broad range of recruitment and players to get used to the shift in in-game balance, in case they would develop (positive or negative) opinions on implemented changes that correspond to the statements of the questionnaire. It was released over the most popular communication channels and platforms as mentioned in Section \ref{Survey} and participants were asked to share it with peers and communities in order to reach as much of the audience as possible, while covering the largest parts of the player expertise spectrum. After completing the quantitative parts about their requirements for and understanding of balance \blue{(cf. Table \ref{tab:SurveyItems})}, they had the opportunity to mention particular builds and professions that needed balancing in either direction, and could comment on their decisions, opinions and mindset through open-ended questions \blue{(cf. Table \ref{tab:QualitativeItems})}. After the collection of survey responses and the compilation of data-driven evidence through in-game combat logs, the outcomes were communicated back to the community in a conservative and as objective as possible manner, which again encouraged discussions and reflections on the notions of balancing.

\subsection{Participants}
\textit{Guild Wars 2's} player base is estimated to exceed 18 million users from which approximately 350,000 players log in and play on a daily basis \footnote{https://mmo-population.com/r/guildwars2}. Even if not all of them participate in the modes of challenging endgame PvE, 
calculating the minimum sample size for this population yields at least 385 responses (assuming a confidence level of 95\% and a margin of error of 5\%) \cite{cochran1977sampling}. After the recruitment period of four weeks, we collected responses from ($n = 680$) players in total. As initially anticipated, these stem from the complete spectrum of experience from 0.5 to 10 years in-game ($M{=}6.6, SD{=}3$), 0 to 10 years in 5-player endgame PvE content ($M{=}3.9, SD{=}2.7$) and 0 to 7 years in 10-player raids ($M{=}2.9, SD{=}2$), where the latter only existed for 7 years at this point of time.

\section{Survey Results}
The following section reveals quantitative outcomes of the balance survey and qualitative comments on the participant's decisions and requirements. Subsequently, specific professions and builds are highlighted that deemed to be imbalanced by the community, before data-driven imbalance metrics are adduced for empirical comparison in the next section. 

\begin{figure}[!h]
    \centering
    \includegraphics[width=\textwidth]{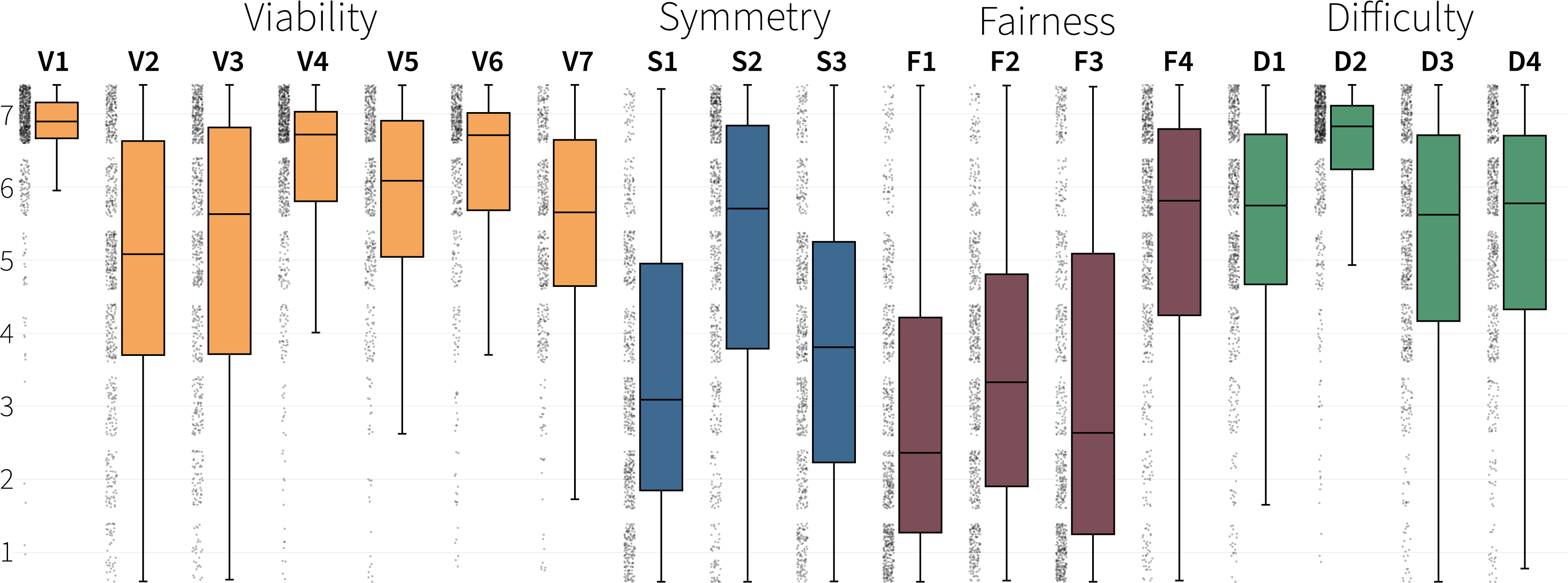}
    \caption{Distribution of survey responses regarding the participants' stances on the balancing concepts \textbf{viability}, \textbf{symmetry}, \textbf{fairness} and \textbf{difficulty}. Answers to the questions of Table \ref{tab:SurveyItems} ranged from 1 ("Strongly Disagree") to 7 ("Strongly Agree"), boxplots indicate means (--), standard deviations (boxes), range (whiskers) and frequencies of an answer (dots).}
    \label{fig:BalanceSurveyBoxplot}
\end{figure}

\subsection{Balance Survey Items}
\label{sec:balanceSurvey}
Figure \ref{fig:BalanceSurveyBoxplot} visualizes the outcome of the community's opinion across the game-specific factors of the different balance concepts (cf. Table \ref{tab:SurveyItems}). Regarding \textbf{viability}, players strongly agree that each of the nine professions should be viable in endgame PvE ($V1: M{=}6.7, SD{=}0.8$), but only slightly when it comes to the 27 specializations ($V2: M{=}4.9, SD{=}1.8$). It would be desirable if every profession has a viable build for the main roles of the game, i.e. power damage, condition damage and support ($V3: M{=}5.1, SD{=}1.9$), but even more important, the roles of power and condition builds should be balanced looking at their viability at the different encounters ($V4: M{=}6.2, SD{=}1.2$). Extreme builds without any convenient utility should be part of the game ($V5: M{=}5.8, SD{=}1.4$) and having these drawbacks should definitely be rewarded in the dps outcome ($V6: M{=}6.1, SD{=}1.3$), while bringing utility to the group should considerably lower the dps potential of a build ($V7: M{=}5.4, SD{=}1.4$).

On the other hand, players rather disagree on desiring a perfect \textbf{symmetry} among the playable options when it comes to dps ($S1: M{=}3.4, SD{=}1.8$). This similarly holds for support builds, as having unique professions to provide certain boons is rather desired for class diversity ($S2: M{=}5.1, SD{=}1.9$), which inherently means the opposite of symmetry. Yet, there are mixed opinions on whether this uniqueness actually makes the formation of groups more limited, which would be unfavorable for balancing ($S3: M{=}3.8, SD{=}1.9$).

When asked about the \textbf{fairness} of choice in playing builds and professions at one's pleasure, players might be potentially concerned if other players would allow their character in a group, even if it provides fewer dps or utility than its alternatives. This is slightly disagreed for both 10-player content ($F1: M{=}2.9, SD{=}1.8$), as well as 5-player content ($F2: M{=}3.4, SD{=}1.8$), the latter being a bit more restrictive on the role coverage. Most build options (the choice of equipment, active skills and passive straits) are heavily influenced by optimized guides written mostly from the experienced speedrun community. This might impact how fair players perceive their own decision making and preferences, yet the community slightly disagrees that this unnecessarily influences fairness of choice ($F3: M{=}3.3, SD{=}2.2$), but rather adjudges that these guides help new players in the first place ($F4: M{=}5.3, SD{=}1.8$).

Finally, the community has a strong opinion towards \textbf{difficulty} and its role in the balance between professions and builds. They welcome that easier options should at least be viable in endgame PvE ($D1: M{=}5.4, SD{=}1.5$), but that higher difficulty should absolutely come along with higher performance ($D2: M{=}6.5, SD{=}1.1$), while failing to execute a more difficult build should equally be punished harder in dps outcome than for easier alternatives ($D3: M{=}5.3, SD{=}1.6$). The same holds for rotations where the damage uptime is harder to realize ($D4: M{=}5.4, SD{=}1.6$).

With respect to the associated game update patch, the community denied to be satisfied with the changes regarding balancing ($M{=}2.2, SD{=}1.3$). \blue{There was no significant correlation between overall in-game experience (or of any game mode) and this satisfaction ($p > 0.05$), nor did the previously asked balancing opinion scales differ between players that focused on raids, fractals or strike missions.}

\subsection{Qualitative Analysis}
\label{sec:qualitativeAnalysis}
Following structured content analysis \cite{mayring2004qualitative}, we \blue{categorized} qualitative responses \blue{deductively (with labels mainly derived from the formerly denoted balance concepts \textbf{viability, symmetry, fairness} and \textbf{difficulty}) \cite{becker2020game} and reflect on} 
the most prevalent opinions \blue{across the open-ended questions (cf. Table \ref{tab:QualitativeItems})}. 
\blue{Two members of the research team carried out this labeling independently, before comparing and discussing their outcomes. In cases of conflict, underlying literature was consulted which in the most cases produced consensus between the annotators eventually (judging from the high inter-rater agreement of Fleiss' $\kappa=0.92$ \cite{fleiss1971measuring}).}

Regarding \textbf{viability}, players highlight that they want all classes to be viable \blue{(36 mentions)}, \textit{``not every profession has to perform equally but every profession should be useful''} (P461) -- as long as this is not \textit{``homogenizing all professions''} (P57), i.e. choices do not become symmetrical. This could be realized by e.g. \textit{``aiming for an average bench of 37k [dps] for most classes [...] with a 2-3k margin in either direction depending on difficulty of class or on the amount of utility they provide [...] This would leave almost everything viable for the majority of groups''} (P66). Apart from this, they state that replacing viable options with stronger new options (``power creep'') is a dangerous step as \textit{``we want horizontal not vertical movement with elite specs. This is exceedingly important to keep content relevant''} (P440).

Perfect \textbf{symmetry} \blue{was strongly opposed (45 mentions) and} reportedly \textit{``destroys the reason why you play a certain spec or even profession. I do not think that a homogeneous cluster of 9 different classes is what GW2 should look forward to''} (P57). The surveyed community rather opposes \textit{``creating a uniformity''} (P75) which would \textit{``lead to a very stale raiding environment''} (P163)
as \textit{``making everyone do the same thing, [...] removing the flavour from the game is not going to be the solution for making the end game more accessible, it's only going to remove the uniqueness and the feeling of every class''} (P300).
This especially holds (but is not constrained to) profession-specific special buffs that \textit{``provided diversity and that will be sorely missed''} (P582), \textit{``[were] more senseful [...] to make professions and specs more unique''} (P57) and removing this factor \textit{``goes against the core of a roleplaying game''} (P486). It has to be remarked though that players do not want these factors of a class seen as mandatory, which would restrict the choice of options again, but \textit{``unique buffs are good as long as they are situational and not required''} (P122).

Participants did not comment extensively on criteria of \textbf{fairness} \blue{(17 mentions)}, but occasionally felt treated unfair when balance decisions went against their own perceptions, e.g. \textit{``I was shocked to see mechanist receiving buffs after it was already one of the best specs in endgame PvE. I would love to see more variety in the game''} (P142). On the contrary, certain voices express that ideal and optimal viabilities are not the primary factor of playing, as \textit{``most of the community prefer playing the class they love and getting the reward, rather than doing a very fast kill with a class they don't enjoy''} (P300) -- which is yet challenged by the emergence of a magnitude of players that do not feel rewarded and \textit{``there would be no reason whatsoever''} (P65) playing a build whose performance is overshadowed by most alternatives.

From the extent of univocal opinions and convictions, \textbf{difficulty} turned out to be the most important factor in estimating the quality of balance in a build, profession or the entire game \blue{(83 mentions)}. Participants agree that a \textit{``low barrier to entry is good [and] after entering, skill expression should be rewarded well''} (P483), \textit{``decently difficult rotations [should lead to] increased DPS, [provided] some low effort average dps classes for newer players [exist]''} (P276) and \textit{``easier builds/rotations for beginners [are] viable, but high end builds should have more complicated rotations that are rewarded with higher damage or utility''} (P297).
Participants made sure that this should not primarily preserve a gap between low- and high-skilled players, but the incentive to self-improve and develop should be significant, as games should \textit{``teach players [...] how to use the combat at higher skill levels''} (P184) and \textit{``bring back the fun in having something to break your mind into a better player optimization world''} (P191). \textit{``It should actually be worth it to learn more complicated builds/rotations and performing them well should have a noticable impact''} (P297) and \textit{``high end difficulty should be kept high to encourage players to learn and reward those who put in the time and effort to understand the game design''} (P321).
This can also be expressed as bringing \textit{''the skill floor down and not the skill ceiling''} (P132), where the skill floor would denote entry levels of performance (``easy-to-learn'') and the skill ceiling stands for ideally ``hard-to-master'' gameplay that produces formidable and rewarding outcomes.
Balancing that fails to acknowledge this requirement \textit{``gives people no reason to learn difficult classes but instead to faceroll with easier classes''} (P184), as \textit{``simplifying all the specs just makes this game boring and unenjoyable''} (P457) and even more critically, player experience and perceived fairness would again be impaired (\textit{``players not wishing to understand the game or work at improving should not be catered to at the expense of players who want to engage with game mechanics''}) (P321).
Eventually, players summarize their requirements on the role of difficulty within balancing (and specifically regarding performance) as \textit{``classes that can only deal dps [should] deal max possible dps, any boon/heal/support/range/cleave/mobility/etc it has reduces this max dps''} (P31) -- or, to quantify this statement -- \textit{``the DPS output of a class should be a function of rotational complexity, boons it can give, self-sustain/squishiness, range/safety, CC contribution, and so on''} (P540).

Apart from the already discussed balancing concepts, players certainly acknowledge that \textit{``only changing values won't cut it''} (P232), so that flair, in-game dynamics and play styles have an equally important impact on players' choices of professions; they desire \textit{``design notes that clearly state reasoning behind changes''} (P393) to follow the intentions of developers and are convinced that player-driven input can positively impact balancing decisions: \textit{``Please [...] consider the opinions of the community before making balancing decisions''} (P383), and \textit{``bring the community back into game''} (P674).

\subsection{Particular Perceptions of Imbalance}
\label{sec:perceptionsOfImbalance}
\begin{figure}[h!]
    \centering
    \begin{subfigure}[t]{0.5\textwidth}
        \centering
        \includegraphics[width=0.9\textwidth]{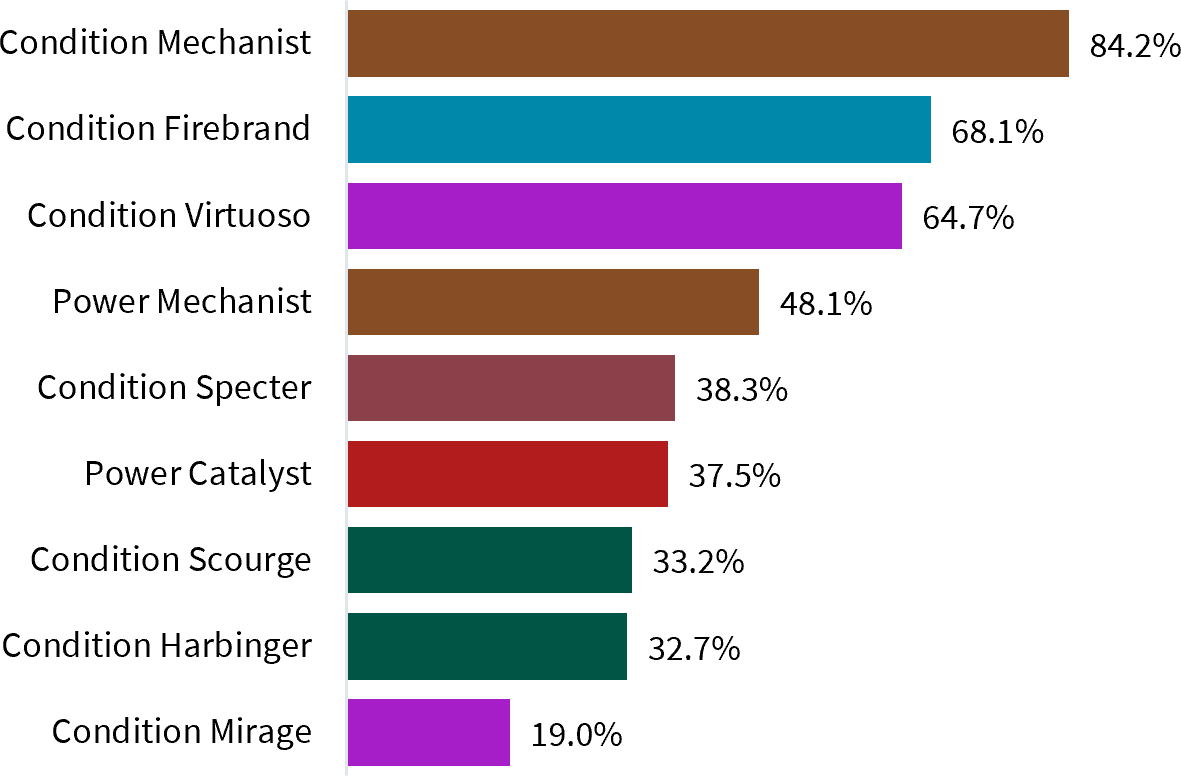}
        \caption{Opinions on builds to be nerfed}
    \label{fig:NerfWant}
    \end{subfigure}%
    ~ 
    \begin{subfigure}[t]{0.5\textwidth}
        \centering
        \includegraphics[width=0.8\textwidth]{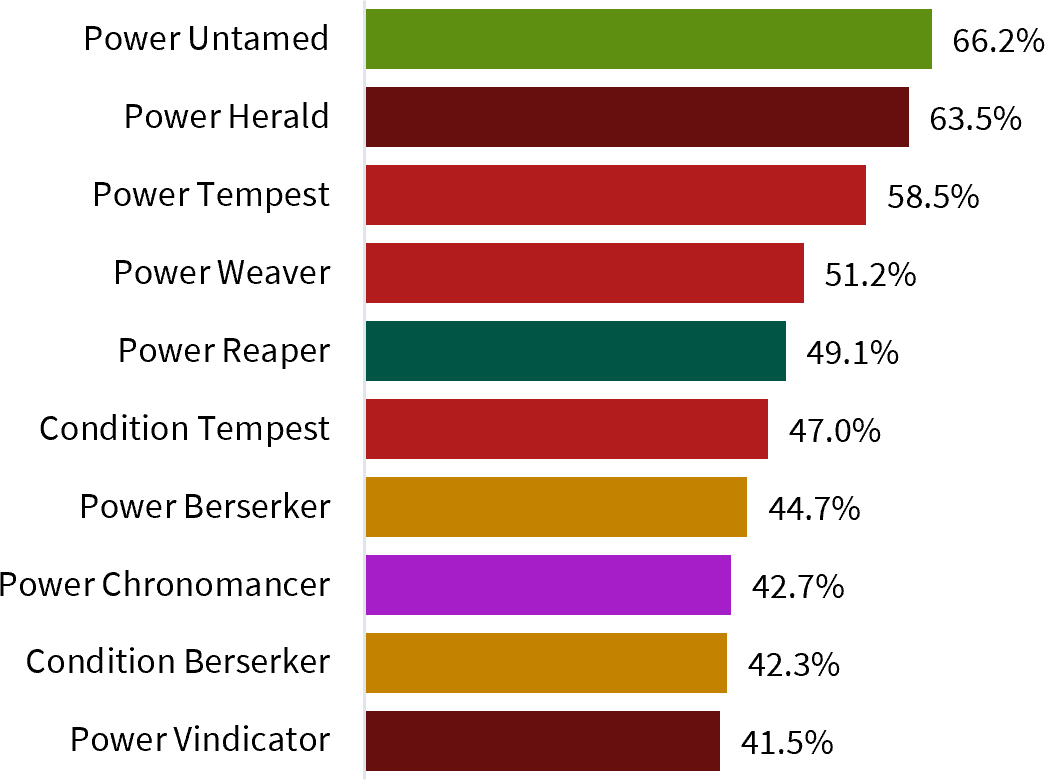}
        \caption{Opinions on builds to be buffed}
    \label{fig:BuffWant}
    \end{subfigure}
    \caption{Popular builds of the game that could benefit from balancing into higher or lower viability or performance, according to the community. Bars and percentages represent the share of participants that indicated their opinion.}
    \label{fig:BuffNerfWant}
\end{figure}

To go deeper into the community's perception of what professions or builds they feel needed a nerf, buff or were quite balanced (before the patch), we additionally asked their opinions so that these could be compared to the actually implemented changes later. Figure \ref{fig:BuffNerfWant} displays the most desired changes of specific builds indicated by the survey participants. Most notably, the general direction suggested for condition classes to be nerfed and power classes to be buffed even before the update patch hit. The only exception was the \textit{Power Mechanist}, which already outshined supporters of a similar role, and \textit{Power Catalyst}, which although rarely played had a drastically high performance potential on the upper end of the skill ceiling. To give some context, this \textit{Catalyst} and the \textit{Condition Mirage} build are the only exceptions among the mentions in Figure \ref{fig:NerfWant} that demand highly difficult and complex gameplay in order to produce decent performances, while all remaining professions in that column had high dps potentials while only requiring minimal effort and player skill.
On the other hand, players desired buffs for professions that are rarely played and/or heavily underperform in comparison to their complexity (cf. Figure \ref{fig:BuffWant}). For the sake of brevity, we exclude players' perceptions on well-balanced classes, as only mechanisms to detect imbalances are discussed in the later stages of the paper.

\section{Data-Driven Foundation}
The former quantitative and qualitative analyses of the opinions of the player community revealed a series of insights that mostly went into a similar direction. 
\blue{
While these could be utilized to inform balance decisions already,}
opinionative findings could be warped or biased by the reached sample, diverging experiences and preferences or the influence the recent game update patch might have had on their perceptions of balance. While we claim that a major shift in game balance does not necessarily change the players' perceptions of balance itself, but rather point out and highlight one's own requirements, beliefs and understandings of balance, we still declare that empirical foundations should be consulted to evaluate and ideally solidify the community's views. To approach this endeavor, we adduce data and tools from the largest analytics platform for \textit{Guild Wars 2} endgame PvE 
\cite{wingman2023}. At the time of evaluation, it comprised ($n_{l}=4,318,009$) recorded atomic boss logs from ($n_{a}=154,145$) unique players and keeps track over the state of balancing and the game in general for the last five years. We deploy measures about the overall usage of different professions throughout the game modes, as well as empirical (dps) performances 
on actual boss encounters across all available builds.

\subsection{Profession Popularity}
\label{sec:professionPopularity}
\begin{figure}[!h]
    \centering
    \includegraphics[width=\textwidth]{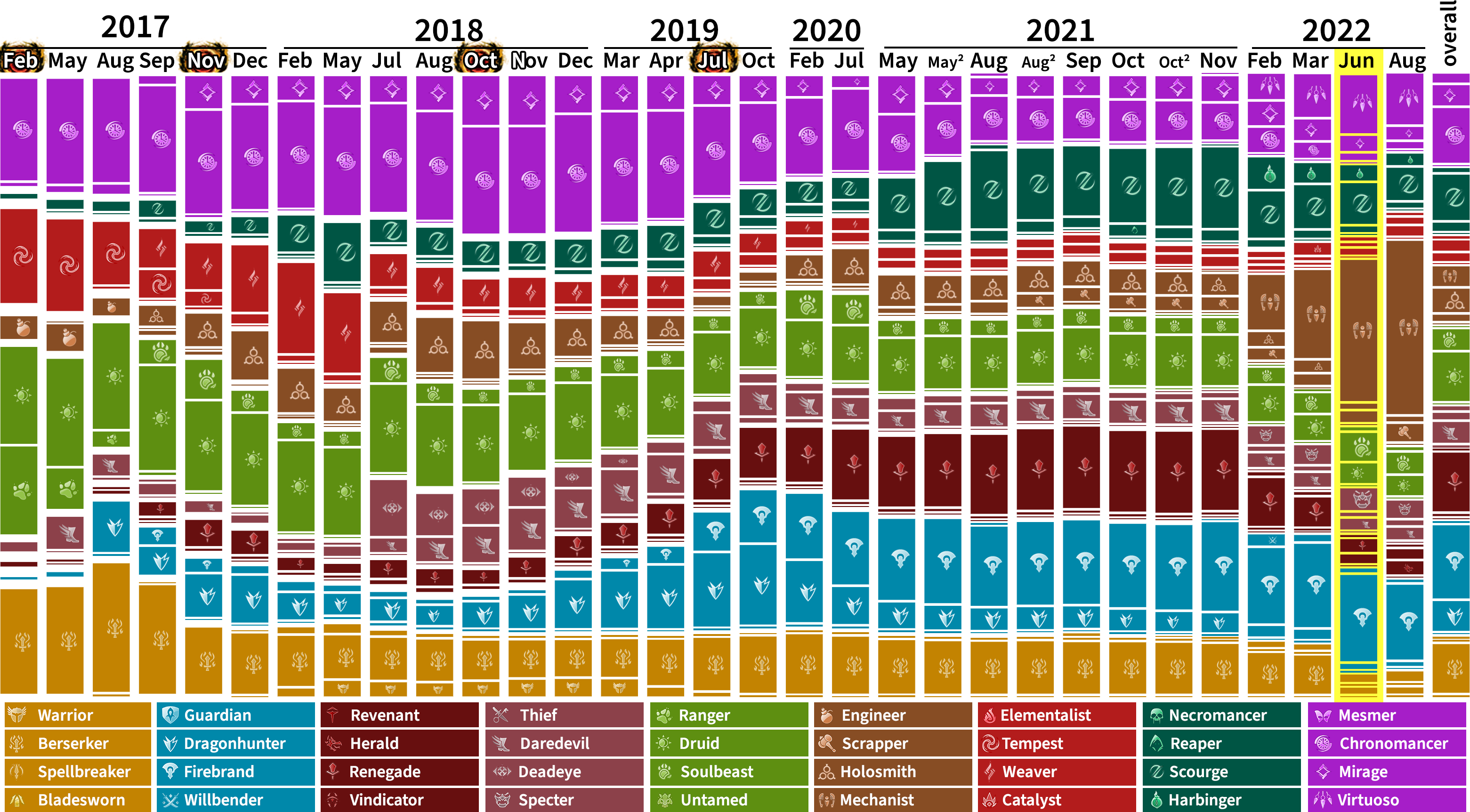}
    \caption{Popularity of the professions and their specializations (\blue{as referenced in the legend below}) used in 10-player \textit{raids} across patch eras from the years 2017 to 2022 (including $7,954,291$ players). In September 2017 and Februar 2022, new expansions were released that added nine additional specializations each. The discussed balance patch is highlighted in yellow, whereas the rightmost bar depicts the overall distribution across all eras. \blue{\classiconOnly{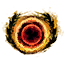} Icons below the patch labels indicate new content releases of raids.}}
    \label{fig:PopularityRaids}
\end{figure}

While not producing detailed assertions on the efficiency, utility, performance or viability of a profession or build, the popularity or usage of a class over longer periods of time can already indicate and immediately highlight trends in balance shifts. Figure \ref{fig:PopularityRaids} visualizes the popularity of each profession and their respective specializations in 10-player \textit{raids} over significant balance patch updates of the years 2017 until 2022. The patch that accompanied the survey of this work is elevated in yellow.

At the time of closing the survey (four weeks after the launch of the June 2022 balance patch), we calculated the change in popularity of the specific specializations. The biggest shifts could be found in the usage of the \classicon{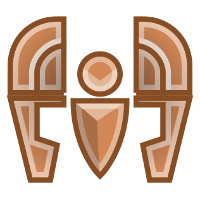} ($+7.1\%$), \classicon{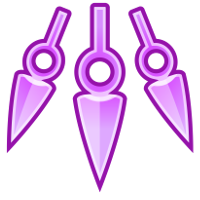} ($+2.6\%$) and \classicon{Soulbeast} ($+1.5\%$), as well as decreasing usage of \classicon{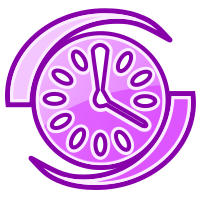}s ($-1.5\%$), \classicon{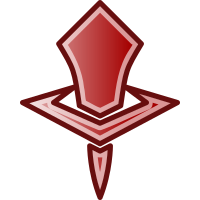}s ($-2.5\%$) and \classicon{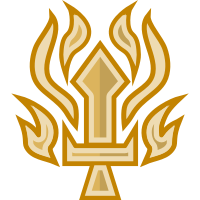}s ($-6\%$). Similar trends with the same affiliated professions could be found for the content of 10-player \textit{strike missions} and 5-player \textit{fractals}. The following update implemented end of August 2022 even steepened this development with almost every third player ($32.7\%$) playing \classicon{Mechanist} since then.

\subsection{Data-driven balance concept assessment}
\label{sec:DataDrivenBalanceConceptAssessment}
While the former section already indicated a \textit{dominant strategy} that threatens to overrule the viability of alternative choices, it lacks expressive power to estimate or explain changes in efficiency or utility. To assess shifts in the performance of particular (damage) builds, we thus add empirical data from the official api of the accompanied analytics platform\footnote{\url{https://gw2wingman.nevermindcreations.de/api}}.
Judging from the qualitative statements about their requirements, players (among other things) uttered 
that balancing should not only revolve around the tip of the iceberg expert players -- but incorporate the proficiency spectrum of the whole player community. For this reason, we delve deeper into the potentials of empirically data-driven analyses and consider performance distributions instead of mere top performances. Returning to the formerly discussed balance concepts, we hypothesize that at least \textbf{symmetry}, \textbf{difficulty} and \textbf{viability} of professions or builds can be measured based on empirically recorded evidence, entailing their respective understandings from the community. This section shortly outlines the theoretical quantification of these balance concepts from damage performance distributions, before we apply them to the community's logs during the study period.

\begin{figure}[h!]
    \centering
    \begin{subfigure}[t]{0.33\textwidth}
        \centering
        \includegraphics[height=0.7\textwidth]{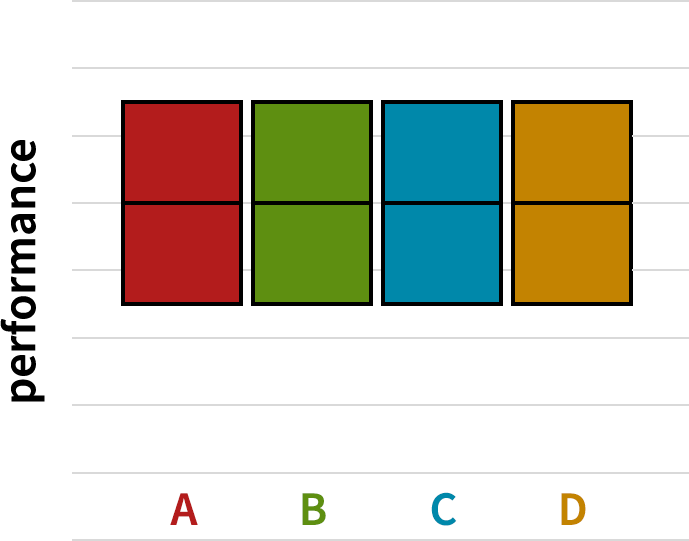}
        \caption{Perfect performance \textbf{symmetry}}
    \label{fig:Symmetry}
    \end{subfigure}%
    ~ 
    \begin{subfigure}[t]{0.33\textwidth}
        \centering
        \includegraphics[height=0.7\textwidth]{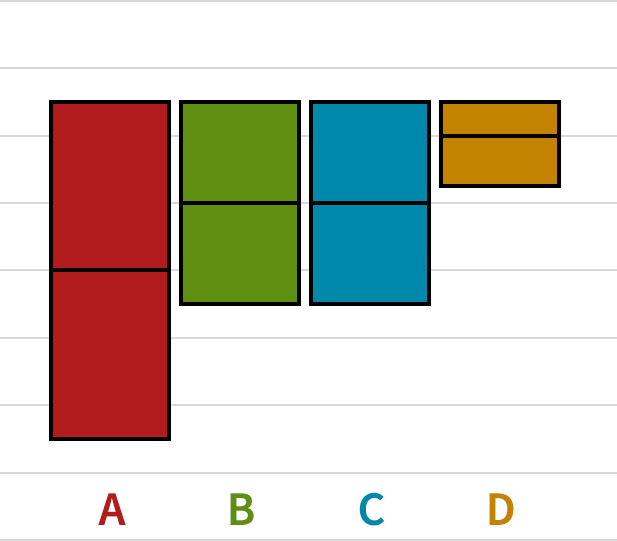}
        \caption{Measurable \textbf{difficulty} differences}
    \label{fig:Difficulty}
    \end{subfigure}
    ~ 
    \begin{subfigure}[t]{0.33\textwidth}
        \centering
        \includegraphics[height=0.7\textwidth]{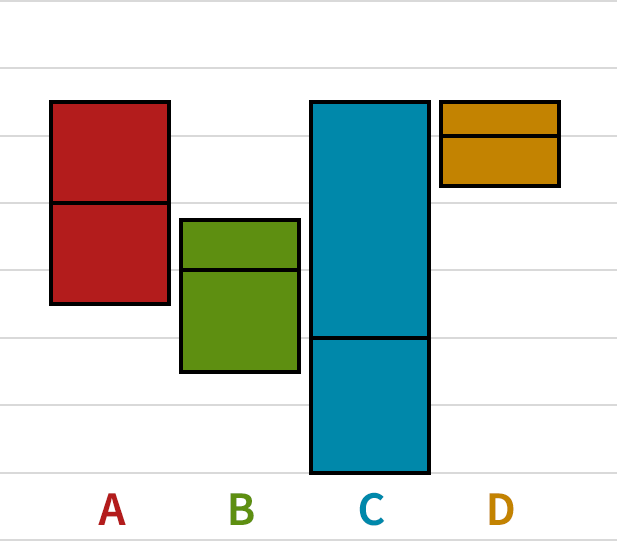}
        \caption{Detectable \textbf{viability} differences}
    \label{fig:Viability}
    \end{subfigure}
    \caption{Empirically quantifiable measures of \textbf{symmetry}, \textbf{difficulty} and \textbf{viability} between four theoretical profession distributions}
    \label{fig:BalanceConceptsDistributions}
\end{figure}

\subsubsection{Theoretical Assessment}
Figure \ref{fig:BalanceConceptsDistributions} visualizes the theoretically quantifiable realizations of these concepts in the data of performance distributions. Following the initial definition of \textbf{symmetry} (i.e. identical expected performance of choices), Figure \ref{fig:Symmetry} depicts perfectly symmetrical performance distributions between the four example options \textbf{A}, \textbf{B}, \textbf{C} and \textbf{D}. Deviations from this symmetry can be quantified by distance metrics as basic as mean squared errors (from assumed equal performance) -- and plotted over time (or balance patch eras) to track the impact of these patches onto the empirical performance symmetry. This similarly holds not only for performance distributions, but also formerly mentioned popularity proportions.

With respect to the \textbf{difficulty} of a specific profession or build, we previously identified that most importantly the proficiency gap in mastering its complexity, as well as the challenge of realistically executing it on a efficient level (influenced by factors of damage uptime and boss-related mechanics), determine its outcoming damage performance. This entails that builds with low complexity pose lower risks of producing worse performances, and also that builds with easy damage uptime produce better performances with a higher probability across the distinct boss encounters and player proficiencies. As exemplified in Figure \ref{fig:Difficulty}, builds with higher difficulty would thus lead to an increased \textit{variance} within performance distributions, i.e. option \textbf{A} depicts example outcomes of a complex profession with hard-to-master damage uptime, while option \textbf{D} represents a simplistic build that is able to consistently deliver its damage across encounters. Even if all of the example distributions are theoretically able to reach the same top values, the impact of difficulty on their realistic performances cannot be neglected.

Assuming that difficulty varies and different builds fluctuate around different median performances, Figure \ref{fig:Viability} showcases differences in \textbf{viability} that can be intuitively visualized and detected. Even if option \textbf{B} is showing significantly lower performances as \textbf{A}, it still should not considered as necessarily unviable, as either some good players of \textbf{B} are still producing better outcomes as some of \textbf{A} or some encounters just favor the usage of \textbf{B} over \textbf{A}, which would open a niche that renders \textbf{B} viable in that case. In contrast, option \textbf{D} strongly \textit{dominates} \textbf{B}, as all of their performances are strictly higher than \textbf{B}, no matter the player proficiency, boss encounter, group constellation or other factors. This essentially prevents it from being a viable choice, as there would always be a choice that makes \textbf{B} a meaningless decision (assumed it does not support the combat in any other way). Despite \textbf{C} having the lowest median and minima, it would not be considered unviable after this conception, as (depending on player skill or situational usage) there are situations it can shine.

There is no quantifiable measure of \textbf{fairness} in the performance data we consulted yet, as players' conceptions of fairness rather involve subjective perceptions not covered in these logs. Yet, if following the initial notions of Becker and Görlich \cite{becker2020game}, fairness can manifest in the probability of succeeding with the option a player prefers, which can be empirically validated by measuring the impact and contribution a build makes on the \textit{success rate} (i.e. proportion of boss kills versus failed attempts) and compared across alternatives. As this work mainly focuses on complying with player-driven understandings, we omit the corresponding analysis for the sake of brevity and focus.

    \begin{figure*}[!h]
        \centering
        \includegraphics[width=\linewidth]{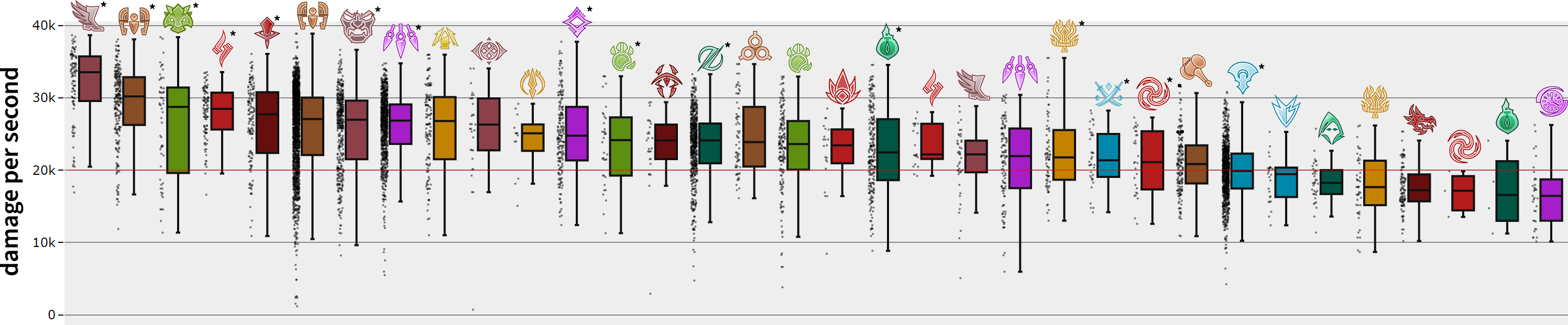}
        \caption{Dps performance distributions of 34 builds over a sample of $1000$ recorded logs of a single boss during the first weeks of the balance patch era in question (including $\sim10,000$ players). Professions are sorted by their median dps and displayed via data point clouds as well as box plots. Condition builds are marked with *, support builds were excluded for the purpose of this visualization.}
        \label{fig:EncounterAnalysis}
    \end{figure*}

\subsubsection{Empirical Assessment}
Bringing these empirical assessments together, we tackle the measurement of the balancing concepts on actual recorded data. As bosses in \textit{Guild Wars 2} are fundamentally different, they afford the usage and preference of various builds and professions, and performance distributions are therefore inherently disparate, these assessments are executed on a per-boss basis and aggregated eventually. Figure \ref{fig:EncounterAnalysis} displays the performance distributions of the most prominent builds for one boss during the weeks the survey was active. With respect to this encounter, we can among other things confirm that the high complexity of the \classiconType{Condition}{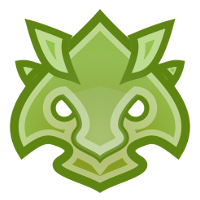} 
expresses in the comparably high variance of performances, indicating \textbf{difficulty} differences. While the best players of this class achieve almost unparalleled values, these records are also reached from a lot larger share of the \classiconType{Condition}{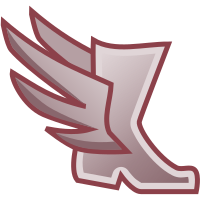} players, suggesting a lower difficulty.
Barring outliers, the missing \textbf{viability} of \classicon{Chronomancer} 
or \classicon{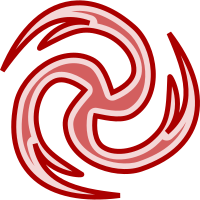} 
also becomes apparent, as more optimal choices such as \classiconType{Condition}{Daredevil} 
or \classiconType{Condition}{Mechanist} 
completely \textit{dominate} these in this context.

When deployed on the full set of the game's 24 \textit{raid}, 9 \textit{fractal} and 10 \textit{strike mission} bosses tracked by our platform \blue{(through $n_{l}=4,318,009$ combat logs)}, this approach ranks \classiconType{Condition}{Untamed}, \classiconType{Condition}{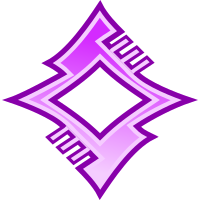} and \classiconType{Power}{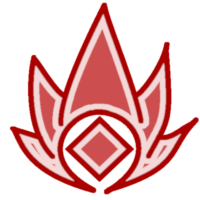} as the currently most \textbf{difficult} (damage) builds of the game, while the \classiconType{Condition}{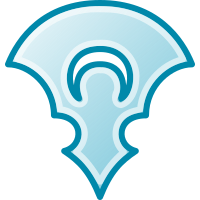}, \classiconType{Condition}{Virtuoso} and \classiconType{Power}{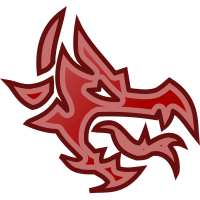} are located on the lower end of the difficulty spectrum. This aligns with subjective difficulty ratings of the particular builds composed by experienced professional players of the game\footnote{https://snowcrows.com/builds}$^{,}$\footnote{https://lucky-noobs.com/builds/condition-daredevil-dps}.

Additionally, we accumulated the occurrences of being \textit{dominated} by another option across all bosses \blue{(within the study period, $n_{l}=146,230$ logs)}. Unsurprisingly, this \textbf{viability} ranking showed most of the game's core professions (without any beneficial specialization) as least viable. After these, \classiconType{Power}{Druid}, \classiconType{Power}{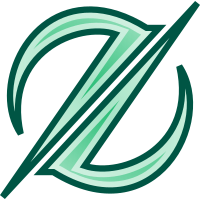} and \classiconType{Condition}{Chronomancer} have shown to be most unviable in the recent endgame PvE of \textit{Guild Wars 2}, which is attributable to the opposite damage type design of these classes.

\section{Discussion}
Balancing video games can follow a multitude of different definitions, approaches and takes. To understand why players are satisfied or frustrated with current states of balance and balancing decisions, we collected notions and understandings of academia and industry, carried out a player-centric survey to assess the game-specific nuances and community-driven requirements based on the former categories, and deployed empirical assessments grounded on an abundance of atomic in-game data logs.
Bringing these worlds together, the player base of \textit{Guild Wars 2} wants all of its nine professions to be \textbf{viable}, with at least some of their currently three choosable specializations, ideally being able to take the role of a (power or condition) damage dealer or support \blue{(questions V1-4, cf. Figure \ref{fig:BalanceSurveyBoxplot})}. This viability can partially be reflected in the popularity, or usage, these professions show in comparison to their alternatives \blue{(cf. Section \ref{sec:professionPopularity})}. Seeing little to no usage of a profession suggests that their viability is threatened, which most commonly correlates with it being not efficient enough, implying a play style that is too complex or unrealistic to execute in actual boss scenarios, or being dominated by alternative choices that are superior in all or most respects. Some voices expressed that players often rather choose their class in endgame PvE on a preference, flair and style basis, but the significant shifts in profession popularity clearly indicate that balancing decisions and the resulting viability of builds are the driving force of what players bring into \textit{raids} (and other content) how frequently. This can partly be explained by the optimal performance these professions can achieve under ideal circumstances, 
but balancing after only top players and performances neglects the vast majority of the player base which should be considered to retain satisfaction throughout the community. Thus, we deployed data-driven assessments of which classes are really lacking viability across the board, based on the performances of hundreds of thousands of players within the weeks of the study period \blue{(cf. Section \ref{sec:DataDrivenBalanceConceptAssessment})}. Admittedly, viability does not stop at afflicting as much damage as possible to the target encounter, but is highly influenced by the amount of \textit{utility} a build can provide that potentially increases the offensive or defensive capabilities of other players and/or leads to increases in the success rate of the combat. Nevertheless, when striving for as fast and clean boss kills as possible (which is the most common take players seek to implement), this \textit{utility} can be either theoretically calculated (by its contribution to the overall group efficiency/dps) or empirically measured (contrasting it to logs without this \textit{utility}). Thus, the discussed viability assessment strategies (based on dps) do not differ systematically from viability based on dps plus the player's contribution on group performance. This then reflects the community's requirements on the equilibrium between ``selfish'' dps builds and those providing extra utility \blue{(questions V5-7)}. 

While players have a high demand for viability across options, they clearly disapprove the idea of \textbf{symmetry} (all professions being equally efficient) \blue{(question S1)}, as this could lead to meaningless decisions, boring and stale gameplay, no room for optimization and the loss of identity. This holds for the damage potentials of the various builds, as they rather like to see spectra of difficulty demands, use cases and applications, but also for the utility builds can provide. Referring to the latter, having unique utility effects is a desirable way of adding identity to classes and diversity to group, as long as it does not limit or constrain the group formation \blue{(questions S2,3)}. 

The perception of \textbf{fairness} within the surveyed community does not completely align with the understandings from literature. Summarized, players want to be able to choose builds and professions after personal preference and not necessarily after ideal or optimized performances of other players, even if the former is likely to be impacted by the latter. In a sense, the fair probability of succeeding in the game \cite{becker2020game} can be loosely transferred to having a just chance to be accepted in other player's groups, and to being able to defeat a boss encounter with the option of one's preference. However, this rather subjective topic presumably rests on the perception of viability within the community and for the case of \textit{Guild Wars 2}, players rather find that the community allows non-optimal builds \blue{(questions F1,2)} and that expectations on single players usually do not have to compete with high-end professional players \blue{(questions F3,4)}.

Judging from both quantitative as well as qualitative responses, \textbf{difficulty} should be one of the most important, if not the major impact factors for balancing. We note that the community's understanding of this connection slightly differs from academic perspectives, as for most related work, difficulty is a parameter \textit{altered by} (automatic) balancing, most often to maximize the player's perception and duration of flow. In this context, difficulty \textit{emerges} from the complexity of the play style of the profession or build, its (ideal) rotation, the hardness of maintaining damage uptime and similar factors. In that respect, the output to balance is rather the outgoing damage performance of a player, which preferably should be consonant with this difficulty to incentivize rewarding experiences. To a certain extent, both definitions still align if a game provides enough choosable options with different degrees of complexity and accordingly compensating damage performances, so that by choosing the most suitable option, players balance themselves into the proper sweet spot between boredom and frustration (flow).
The community desires accessible gameplay for players throughout the entire proficiency spectrum \blue{(question D1)}, but easier builds should be outperformed by more complex ones \blue{(question D2)} if players executing those are actually capable of fulfilling the higher demands of difficulty and realizability \blue{(questions D3,4)}. Detecting differences in this conception of difficulty of a class turned out to be suitable when quantifying the variance across performance distributions of single builds \blue{(cf. Section \ref{sec:DataDrivenBalanceConceptAssessment})}. This assessment is based on the observation and statements about inefficient players producing lower performances when playing more complex builds, compared to executing more forgiving and easy rotations. 

Ultimately, the major balance patch released in June 2022 was primarily perceived as a deterioration of the state of balancing, judging from the majority (83\%) of negative satisfaction responses to this survey as well as from the overall reception displayed in the previously mentioned communication channels. Interpreting the qualitative and quantitative answers from this community sample and contextualizing these with the introduced changes, the balance patch contradicted the requirements of the player base in multiple ways. This can be attributed to a shift in \textbf{symmetry}, as effects that made a number of classes unique were removed, leading to less diversity in group formation, which is not what players were looking forward to. The same change also impacted a series of power classes higher than condition builds, which significantly decreased their \textbf{viability} and performance outcome distributions, up to the point where (both damage and support builds) were completely dominated by other choices. On top of that, opinions on builds that should have been nerfed were either not fulfilled or even reversed, e.g. the unwelcome buff of the \classiconType{Power}{Mechanist} or the \classiconType{Power}{Chronomancer} nerf that went against their expectations \blue{(cf. Section \ref{sec:perceptionsOfImbalance})}.
The biggest perception of \textit{unfairness} in balancing was attributed to the mismatch of \textit{difficulty} and outcome. This becomes apparent when observing the popularity across professions (cf. Figure \ref{fig:PopularityRaids}) in which the \classicon{Mechanist}, \classicon{Firebrand} and \classicon{Virtuoso} alone make up the absolute majority of the used options -- consistently these builds who are easy to play while afflicting decent damage with high uptime and/or providing massive group utility. This is not only reflected in their popularity, but also in the empirical performance assessment. 
In the meantime, more difficult professions lost usage as under realistic circumstances, they could no longer compete with performances of simplistic builds and rewarding playing experiences diminished.

Nevertheless, the abundance of qualitative and quantitative feedback from the player community motivated the developers to increase the viability of a series of neglected or negatively affected builds in promptly following balancing updates. This again emphasized the vital importance --  but also their validity and appropriateness -- of the opinions and requirements of a game's community. Throughout this work, we compiled the advantages and possibilities of player-driven game balancing from related academic work and qualitative as well as quantitative responses of a player base, as long as they are backed up from empirical evidence data. 
In immediately succeeding work, we utilized the now extracted insights and measures for the construction of a tool that seeks to overcome occurrences of imbalance \cite{wingmanBalancingTool2023}.

\blue{
\section{Synopsis \& Generalization}
In the following, we will reiterate the methodology of connecting player-driven surveys, data-driven assessments and deriving an instrument towards balancing video game elements, while answering the priorly posed research questions with the help of a case study targeting balance in \textit{Guild Wars 2}.

Balancing (online) video games adjusts different playable elements for equality, viability, popularity or other factors.
In related scientific literature, myriads of (even conflicting) definitions and conceptions of balance and balancing exist \cite{becker2020game}, every industrial game company decides for their own personal strategy \cite{scacchi2017practices} and above that, balance perceptions of the eventual player bases can deviate from these as well \cite{schreiber2010game}. 
This disagreement frequently backfires when balance perceptions of developers do not align with opinions and requirements of the players, which harms player experience, satisfaction, retention and critical as well as economical success of games.
Thus, we approach the answering of \textbf{(RQ1)}: ``With so many conflicting theoretical definitions of balancing, how can a game understand and cater to the requirements of its players?'' by a mixed-methods survey grounded in concepts of scientific literature. As the game elements, mechanics and dynamics essentially predetermine which balance concepts are important for a game, we first identified these crucial features with the help of a subset of experienced players (in the case of \textit{Guild Wars 2}, this resulted in \textbf{viability}, \textbf{symmetry}, \textbf{fairness} and \textbf{difficulty}, cf. Section \ref{sec:construction}). The next step connected these higher-level concepts to in-game terms and ideas that players can understand (cf. Table \ref{tab:SurveyItems}). Together with qualitative assessments, surveys of these can give in-depth insights about how and why a target player base demands, rejects or is indifferent to specified balance concepts (cf. Sections \ref{sec:balanceSurvey}, \ref{sec:qualitativeAnalysis}). Regarding our case study, a representative sample of the \textit{Guild Wars 2} community deemed multiple \textbf{viability} criteria as important, but strongly opposed that every playable class should produce \textbf{symmetrical} performance outputs or follow similar designs.
This equally bears implications for recent scientific work that explicitly balanced towards symmetrical outcomes in the same or similar game genres \cite{pfau2022dungeonsII}.
Above that, in contrast to academic literature on \textbf{difficulty} balancing \cite{andrade2006dynamic,tijs2008dynamic,volz2016demonstrating}, we observed a clear trend in the community that sees difficulty not as the variable to adjust. Rather, they desire a variety of playable options that differ in difficulty -- while the adjusted metric is rather the damage potential (and variance) it can offer, based on that difficulty.
Details of these insights, paired with specific perceptions of imbalanced elements (cf. Section \ref{sec:perceptionsOfImbalance}) thus reveal requirements that do not implicitly follow from literature (or even contradict those), which can help developers to understand the needs of their player base and researchers to extend existing notions of balance.

Nevertheless, one of the biggest drawbacks of subjective or opinionative assessments remains in the fact that player perceptions can easily deviate from actual circumstances in the game. This can happen due to unawareness of the big picture, inexperience of the game, effects of echo chambers when steadily playing with the same set of people and obviously because player-accessible higher- and lower-level analytics are not the standard, but rather uncommon for most games \cite{wallner2018automatic}. 
With this in mind, we want to answer \textbf{(RQ2)}: ``How can data-driven analytics help grounding the objectivity of this pool of opinions?'' twofold. First, we present how (subjective) player-driven insights produced by the previous step can be consolidated by means of (objective) data-driven techniques. This can be seen in the changes within popularity of in-game elements (cf. Section \ref{sec:professionPopularity}) that go along with players' perceptions of reduced viability. When locating dominant strategies moreover, relying on theoretical (or ideal) peak performances 
is a popular method that can give rough first insights, but yet does not always translate to realistic performances in actual combats. 
More importantly, empirical balance assessments should go even beyond that and incorporate \textit{distributions} of performances which can reflect measures of \textbf{difficulty} and \textbf{symmetry} from the data (cf. Section \ref{sec:DataDrivenBalanceConceptAssessment}). If these align with the former subjective outcomes, they arguably support prior insights with empirical evidence, fortifying the conclusions for suitable balancing decisions. 
Second, we aim to support answering this research question by reducing the gap between subjectivity and objectivity of players, so that requirements and opinions follow informed understandings instead of instinctive estimates. To assess how to educate players or entire communities, we introduce an empirical analytics tool for \textit{Guild Wars 2} that follows player-driven design and embed the balancing assessments of this work into it \cite{wingman2023}. This succeeds (position) papers that call for further research on approaches to player-centric analytics for the benefit of both players and science \cite{kleinman2021using,wallner2021players}.

Having subjective as well as objective measures of in-game imbalances and well-balanced target states, we evaluate the integration of both into an interactive instrument in parallel work \cite{wingmanBalancingTool2023}.
While the final approach of assessing a player's opinion of balance (based only on one visualization about performance distributions) is still shallow and an educated perception of balance arguably longs for experiencing the state of the game by playing it a fair amount of time, we still claim to support the assessment and regulation of balance, and strive to evaluate the impact of player-driven balance implementations in live patches. In order to showcase the applicability of this endeavor and to give a hands-on example that player communities are interested in and capable of being incorporated in balancing, we utilized \textit{Guild Wars 2} as a fitting, popular and contemporary use case. Yet, the proposed methodology would certainly also work for instanced (group) PvE content in general, such as raids, dungeons, trials or ultimate encounters in \textit{World of Warcraft}\cite{WoW}, \textit{The Elder Scrolls Online}\cite{ESO} or \textit{Final Fantasy XIV}\cite{FFXIV}. 
The underlying balance concepts (not limited to viability, symmetry, difficulty and fairness) and the aggregation of a community's performance requirements however are arguably generalizable and similarly applicable for single-player or competitive PvP settings such as in balancing champions of Multiplayer Online Battle Arena (MOBA) games (e.g. \textit{League of Legends} \cite{LoL} or \textit{Dota 2} \cite{Dota2}) or class-based first-person shooters (such as \textit{Valorant} \cite{Valorant} or \textit{Borderlands 3} \cite{Borderlands3}) -- which yields great potential for player-driven balancing.
}

\section{Limitations \& Future Work}
The endeavor of balancing, especially with the multitude of playable options, classes and builds of modern online games, is as complicated as finding a convincing definition of balance in the first place. Thus, the presented work undergoes a number of limitations, partly caused by the disparity of the conceptions and partly because of realistic restrictions within the methodology.
Finding balanced configurations in a popular world-class video game that would suit requirements of hundreds of thousands to millions of active players without disappointing anyone is arguably impossible. 
To this respect, we could only reach a comparably minor part of the entire audience of players, even if the sample size is large enough to enable drawing conclusions. In order to accumulate this sample, we broadcasted this survey through the most popular communication channels (outside the game) and made sure to reach players on the full spectrum of game experience, but admit that those players interested in balancing (and in expressing their opinion) were more likely to respond to this survey and might have skewed the results.
The topicality of the survey, as directly accompanying a major balance patch, might have activated more players that are unsatisfied with the changes and thus introduced a further bias. However, the gist of this work focuses on the underlying and basic understandings of balancing itself, which should not be systematically changed, but rather clarified within players by conceiving contemporary examples of what should or should not happen in proper balancing procedures. Findings related to the very patch are somewhat discussed, but placed back in favor for implications of (player-driven) balancing in general.

Developers are moreover responsible to push novelty, enjoyment and shifts out of rigidly stuck constellations, in order to keep their game innovative, interesting and economically competitive. We neglected factors of flair, intrinsic motivation of play styles and further variables not inherently related to performance for now, but acknowledge that player choices and preferences are not completely rational (with respect to efficiency and strategy optimization). The significant changes in profession popularity yet indicate that power and viability of choices do have a drastical impact on what players play in endgame PvE.

In their meta review on the definitions of balancing, Becker and Görlich list even further possible concepts appearing in related work, such as chance, (in)transitivity, positive/negative feedback, economies, costs, rewards or static versus dynamic balancing. While some of them turned out to be inapplicable to the domain of endgame PvE in \textit{Guild Wars 2}, certain factors such as positive and negative feedback can be identified when attributed to the success of playing out an ideal rotation. Future work will look into the intricacies of single builds and rotations to pinpoint the impact of these points onto balancing perceptions instead of involving them simply inside of difficulty.

The biggest critique against utilizing player-driven balancing opinions came from some players themselves that proclaimed that \textit{``players do not know what is good for them anyway''}. While this might hold to a certain extent and the driving mechanical, dynamical and design decisions should undoubtedly stem from the mindsets of developers, this work counted on the wisdom of crowds of a community, which 
\blue{(when validated with data-driven evidence)}
produces reasonable and thoughtful insights - often in accord with parallel scientific research. Eventually, balance decisions impact player experience, satisfaction and the critical and economical success of a game over longer terms, so tailoring it to the needs of the actual audience cannot be evaluated enough.

For future work, we mainly seek to extend and unravel the nuances of play styles, builds and implications for balancing. These will be incorporated in the lastly introduced player-driven balancing tool \cite{wingmanBalancingTool2023} to portray ideal perceptions of balance versus current empirical constellations down to the lowest possible detail. 
\blue{Even though we evaluated this tool in terms of balance perceptions regarding theoretical aggregated configurations, the implementation of such produced balancing decisions into actual gameplay and evaluation of the subsequent player experiences is an important open endeavor. In pursuing this path, we strive to tighten the connection between players and developers -- as well as between industry and academia.}

\section{Conclusion}
Balancing choosable or even customizable options is an interminable, controversial and considerable process for a game, its developers and players alike. Along with the introduction of new content and fixes for bugs, balancing changes are one of the major causes of update patches for online games, display never-ending experience optimization problems and highly impact how players play a game altogether.
Scientific efforts to study balance are divided into several understandings of the term, even partly conflicting. When it comes to this definition of adjusting for equated or appropriate adjustments of in-game options, related work is largely under-investigated and mostly regards simulation- or computation-based approaches to even out viability across choices. The role, perception and requirements of the player or even the actual player community have not been considered in academia so far, despite bearing considerable implications for games user research, game design and human-computer interaction in general.
For these reasons, we aggregated notions of balance from adjacent research and industrial perceptions, recruited ($n=680$) players of the MMORPG \textit{Guild Wars 2} (as a popular representative of online games undergoing constant rebalancing) and refined a game-specific understanding through quantitative as well as qualitative survey items. Analyzing and interpreting the players' requirements, opinions and mindsets explained their reactions on recently implemented balance changes, enabled collateral data-driven assessments, paved the way for finding community agreements on balancing through an interactive democratic tool and entails (or approves) the following implications that likely hold for comparable games and larger concepts of balancing:
\begin{itemize}
    \item Players crave diversity of in-game choices and a high viability across these choices - Balancing should identify and address dominating choices that render alternatives irrelevant.
    \item Perfect symmetry between the outcome of choices would implicitly entail balanced viability, but this risks leaving decisions meaningless - so different options should display different strengths, weaknesses, use cases and challenges.
    \item Improper balance can evoke feelings of unfairness when common factors such as flair or preference influence players' decisions - this might lead to decreasing satisfaction or potentially even churn.
    \item While most academic approaches manipulate difficulty to balance a player's flow state, players' perceptions of difficult choices in online games strongly presuppose rewarding incentives - this does not have to be extrinsic, but should lead to an increased (perception of) efficiency and/or competence.
    \item In order to optimize satisfaction and experience, players desire to be part of the balancing process - investigating and quantifying balance requirements in this player-driven process elevates the very player's role and trailblazes novel approaches of engaging, binding and tailored games.
    
\end{itemize}

\section*{Acknowledgments}
We thank all participants of the survey for their contribution on player-driven balancing analytics, as well as all 
users of our platform for sharing their extensive playing histories, constant feedback and ongoing discussions. We moreover are grateful for ArenaNet for developing \textit{Guild Wars 2}.
Guild Wars 2 and all associated logos, designs, and composite marks are trademarks or registered trademarks of NCSOFT Corporation or ArenaNet, LLC, respectively. ©2021 NCSOFT Corporation, ©2021 ArenaNet, LLC. All rights reserved.


\bibliographystyle{ACM-Reference-Format}
\bibliography{sample-base}

\appendix
\section{Game Terminology}
\label{appendix}
\begin{longtable}{r|l}
    
    \multirow{2}{*}{atomic}	& \textit{(as in atomic actions)}: Logs of the game utilized in the data-driven evaluation part are recorded on the\\
    &lowest-level possible, i.e. down to \textit{skill} usage and character movement on a frame-by-frame logging basis.	\\
    \rule{0pt}{4ex}
    
    boons	& Temporary positive effects that increase character \textit{stats} or yield \textit{utility}, most often provided by \textit{support builds}. \\
    \rule{0pt}{4ex}
    
    \multirow{2}{*}{buff}	& Increase in damage, \textit{utility} or general viability of a \textit{profession} or \textit{build} caused by balancing adjustments.\\
    &Opposed to \textit{nerf}.	\\
    \rule{0pt}{4ex}
    
    \multirow{2}{*}{build}	& The customizable configuration of equipment, \textit{skills} and \textit{traits} of a \textit{profession}. Most builds target the optimization\\
    &of \textit{power} or \textit{condition} damage output, maximizing \textit{utility} or hybrid versions of these.	\\
    \rule{0pt}{4ex}
    
    class	& The overarching archetype for each character, referred to as \textit{profession} in \textit{Guild Wars 2}. \\
    \rule{0pt}{4ex}
    
    \multirow{2}{*}{condition}	& Temporary negative effects that deal damage over time on an enemy or weaken their \textit{stats}.\\
    &Some \textit{builds} are optimized to deal condition damage in contrast to direct \textit{power} damage.	\\
    \rule{0pt}{4ex}
    
    \multirow{3}{*}{damage uptime}	& The ability of a \textit{build} (or player) to consistently deliver damage, mainly influenced by factors such as survivability, \\
    &range of \textit{skills}, freedom of movement while executing \textit{skills}, adaptability of the ideal \textit{rotation} to live combat, and\\
    &dependence on other factors such as the size of the enemy's hitbox, attack delay or movement patterns. 	\\
    \rule{0pt}{4ex}
    
    dps	& \textit{(damage per second)}: The theoretical or empirical damage a \textit{build} or \textit{player} afflicts onto their target(s). \\
    \rule{0pt}{4ex}
    
    \multirow{4}{*}{endgame} & In \textit{Guild Wars 2}, \textit{PvE} endgame content is mainly carried out in instanced dungeons for five players \textit{(fractals of the}\\
    &\textit{mists)} or ten players \textit{(raids and strike missions)}. As it features no power creep or item spiral and most players\\
    &follow \textit{builds} and \textit{rotations} from community guides, combat logs for single bosses are highly comparable between\\
    &groups and players and differences in efficiency are mainly attributed to the proficiency of players.	\\
    \rule{0pt}{4ex}
    
    \multirow{3}{*}{log} & For the platform used in this work, single bosses or encounters are recorded in \textit{atomic} detail, representing the\\
    &full combat replay and \textit{dps}, heal, \textit{boon}, \textit{condition} among other statistics at every single point in time for up to\\
    &ten players \cite{wingman2023}. \\
    \\
    \rule{0pt}{4ex}
    
    \multirow{2}{*}{nerf}	& Decrease in damage, \textit{utility} or general viability of a \textit{profession} or \textit{build} caused by balancing adjustments.\\ &Opposed to \textit{buff}. \\
    \rule{0pt}{4ex}
    
    performance	& The quantified outcome of a player at a given situation, e.g. for one boss fight. Mostly expressed as \textit{dps} values.	\\
    \rule{0pt}{4ex}
    
    \multirow{2}{*}{power}	& \textit{(as in power damage)}: Direct damage as opposed to \textit{condition} damage (damage over time). Some \textit{builds} are\\
    &optimized to deal power damage.	\\
    \rule{0pt}{4ex}
    
    \multirow{2}{*}{profession}	& \textit{Guild Wars 2's} notion of character \textit{classes}. It features nine core professions that can be extended with one of\\
    &three \textit{specializations} each for more \textit{build} diversity. \\
    \rule{0pt}{4ex}
    
    \multirow{2}{*}{PvE}	& \textit{(Player versus Environment)}: \textit{Guild Wars 2} features single- and collaborative multi-player modes.\\
    &This work focuses on the group-based \textit{endgame} content.	\\
    \rule{0pt}{4ex}
    
    \multirow{2}{*}{PvP}	& \textit{(Player versus Player)}: \textit{Guild Wars 2} features small- and large-scale competitive PvP modes, yet to keep the\\
    &assessment as concise as possible, we focused on balancing \textit{PvE} in this work.\\
    \rule{0pt}{4ex}
    
    rotation	& The sequence of \textit{skills} players execute, often looping for ideal rotations (optimizing \textit{dps}) within a \textit{build}.	\\
    \rule{0pt}{4ex}
    
    \multirow{2}{*}{skill}	& \textit{(as in executable skills)}: Single actions players activate to deal damage and/or provide \textit{utility} by pressing\\
    &the corresponding button.	\\
    \rule{0pt}{4ex}
    
    \multirow{2}{*}{skill}	& \textit{(as in player skill)}: The proficiency of a player (on a specific \textit{build} or \textit{profession}), quantifiable in the amounts\\
    &of \textit{dps} or \textit{utility} they can provide.\\
    \rule{0pt}{4ex}
    
    stats   & In-game character attributes that influence damage, \textit{utility} or survivability potentials of a \textit{build}. \\
    \rule{0pt}{4ex}
    
    \multirow{2}{*}{specialization}	& \textit{Professions} can be added one of three specializations (27 in total) that affect the mechanics, \\
    &damage and/or \textit{utility} potentials of a \textit{build}.	\\
    \rule{0pt}{4ex}
    
    support	& As opposed to \textit{power} or \textit{condition} damage \textit{builds}, support roles mainly provide \textit{utility}. \\
    \rule{0pt}{4ex}
    
    \multirow{2}{*}{trait}	& Customizable passive perks of a \textit{build} that mostly increase damage or \textit{utility} potentials, as opposed to\\
    &active customizations (\textit{skills}). \\
    \rule{0pt}{4ex}
    
    \multirow{2}{*}{utility}	& Beneficial value a \textit{build} can provide for itself and/or other players in the group (apart from \textit{dps}), such as \textit{heal},\\
    &\textit{buffs}, movement \textit{skills}, \textit{conditions}, crowd-control or the ability to resurrect fallen players.\\

    \caption{Terms especially used in \textit{Guild Wars 2}, the genre of MMORPGs or their analytics}
    \label{tab:GameTerminology}
    \end{longtable}

\end{document}